# 1-bit Compressive Data Gathering for Wireless Sensor Networks

Jiping Xiong, Member, IEEE, Qinghua Tang, and Jian Zhao

*Abstract*—Compressive sensing (CS) has been widely used for the data gathering in wireless sensor networks for the purpose of reducing the communication overhead recent years. In this paper, we first show that with simple modification, 1-bit compressive sensing can also been used for the data gathering in wireless sensor networks to further reduce the communication overhead. We also propose a novel blind 1-bit CS reconstruction algorithm which outperforms other state of the art blind 1-bit CS reconstruction algorithms. Experimental results on real sensor datasets demonstrate the efficiency of our method.

*Index Terms*—Data Gathering, Wireless senor Networks, compressive sensing, 1-bit Compressive Sensing

## I. Introduction

IN wireless sensor networks, the sensor readings collected from natural phenomenon usually have inter-spatial and intra-temporal correlation characteristics [1], which means those readings are compressible in some transform domain, such as wavelet transform or discrete cosine transform etc. Depending on the correlations, compressive sensing [2-3] has been widely used to compress sensor readings in recent years.

Wang et al [4] first exploited the inter-spatial correlation of readings between sensor nodes in large scale wireless sensor networks. They proposed a sparse random projections method in a distributed way. In this method, sensor nodes are classified to encoding sensor node and data sensor node. Encoding sensor nodes sum up and store all the weighted readings sent from data sensor nodes, and then base station or collector can reconstruct all readings from querying only small portion of the encoding sensor nodes. The drawback of this method is the high communication overhead between sensor nodes to generate the projections.

In [5], the authors proposed an energy efficient Compressive Data Gathering (CDG) method in large scale multi-hop WSNs. First of all, a spanning tree rooted at a base station will be generated, and the spanning tree contains all the sensor nodes. Second, each leaf node on the spanning tree sends a data vector with length m generated from one reading to its parent node along the spanning tree (given by a m by n random projection matrix, m<n), after receiving all the vectors from all its children nodes, the intermediate nodes sum the vectors up and then send the result vector with length m to its own parent node. In the end, the base station can reconstruct all the original readings from the final vector with m measurement values. The authors further reduced the communication overhead by using a sparse projection matrix instead of random projection matrix [6]. Similar with CDG scheme, Luo et al [7,8] proposed a hybrid CS method to reduce the number of readings each node(especially leaf node) needs to send. Actually, in hybrid CS method, each leaf node only sends its own readings instead of m weighted measurement to its parent node.

Despite the energy efficiency manner, the CDG based methods which utilize inter-spatial correlation at least suffer from following issues: 1) the spanning tree needs to be regenerated even if there is only a single node leaves or joins, thus it results in a very high communication cost; 2) accurate time synchronization is required between sensor nodes in WSN which is not easy to setup and maintain.

In [9], the authors exploited the intra-temporal correlation for WSNs, and proposed methods to make sensor data sparser for CS based methods.

Having observed that the sparsity of readings may vary at different time period, Wang et al [10] suggested using intelligent compressive sensing to adaptively adjust the number of projections each sensor node need to send to satisfy the successfully reconstruction requirement of CS algorithms.

Carlo et al [11] evaluated the computation overhead of generating different projection matrices, and concluded that sparse binary matrix has the lowest generating computation overhead. Also, the authors successfully used Distributed Compressive Sensing [12] to reconstruct sensor readings from projections of multiple sensor nodes.

In [13-16], matrix completion technology [17], which is the extension of compressive sensing, has been used in WSNs to reduce the sampling energy cost. The basic idea is that instead of using uniform sampling, each sensor node randomly samples readings under a certain probability, and sends the incomplete readings to base station. Base station forms an incomplete matrix combining from all the incomplete readings from sensor nodes. Due to the sparsity of readings, the incomplete matrix can be reconstructed using matrix completion theory. The main

This work was supported in part by the Opening Fund of Top Key Discipline of Computer Software and Theory in Zhejiang Provincial Colleges at Zhejiang Normal University.
Jiping Xiong, Qinghua Tang and Jian Zhao are with the College of Mathematics, Physics and Information Engineering, Zhejiang Normal University (corresponding e-mail: jpxiong@ieee.org).

drawback of this kind of methods is that they can't handle abnormal readings.

More recently, 1-bit CS has attracted more and more attention for the ability to balance the cost of more refined quantization and the cost of additional measurements [18]. In this paper, we focus on intra-temporal correlation and first apply 1-bit CS to data gathering in WSNs.

The remaining of this paper is organized as follows. In section II, we first give brief description of 1-bit CS and then propose our framework and algorithm. In section III, we evaluate the efficiency of our method using real WSN datasets. At the end, we conclude the paper and present future research work.

## II. PROPOSED FRAMEWORK

### A. 1-bit CS

Instead of transmitting projection values directly, 1-bit CS only sends the sign of projection values, such as

$$b = sign(\Phi s) \quad (1)$$

where sign is a sign function, such that $sign(y) = \begin{cases} 1, y \geq 0. \\ -1, y < 0. \end{cases}$, $\Phi$ is a given $m \times n$ Gaussian random matrix, $s \in R^n$ has only k ($k << n$) non-zero elements and $b \in \{1,-1\}^m$. Now the problem is trying to recover sparse signal $s$ from sign measurement $b$. Boufounos and Baraniuk [18] first showed that we can accurately reconstruct normalized $\hat{s}$, through

$$\hat{s} = \arg\min_{s} \|s\|_1, \quad s.t.\ B\Phi s \geq 0\ and\ \|s\|_2 = 1 \quad (2)$$

where $\|\cdot\|_1$ and $\|\cdot\|_2$ denote the $l_1$-norm and $l_2$-norm of a vector respectively, and B := diag(b) is a $m \times m$ diagonal matrix whose diagonal elements are from the element of vector b. Also, the constraint $B\Phi s \geq 0$ in model (2) imposes the consistency requirement between the 1-bit measurements and the solution, and the constraint $\|s\|_2 = 1$ limits the solution to a unit $l_2$ sphere which also means the estimated $\hat{s}$ loses the scale information.

In [18], the authors relaxed the constraint and used Fixed-Point Continuation(FPC) algorithm solving below alternative model

$$\hat{s} = \arg\min_{s} \|s\|_1 + l\sum_{i} f((B\Phi s)_i) \quad s.t.\ \|s\|_2 = 1 \quad (3)$$

where $l$ is a tuning parameter and

$$f(x) = \begin{cases} \frac{x^2}{2}, x < 0. \\ 0, x \geq 0. \end{cases} \quad (4)$$

BIHT[19] currently was one of the fastest and accurate reconstruction algorithm which solves following model

$$\hat{s} = \arg\min_{s} \sum_{i} f((B\Phi s)_i) \quad s.t.\ \|s\|_0 \leq k\ and\ \|s\|_2 = 1 \quad (5)$$

But model (5) needs the sparse level k as the prior information which is not practical in WSNs.

We classify those kinds of reconstruction algorithms without the prior requirement of sparsity level as blind 1-bit CS.

In [20], the authors proposed a efficient blind 1-bit CS method which was a restricted step shrinkage(RSS) algorithm to solve model (2). This algorithm used trust-region methods for non-convex optimization on a unit $l_2$ sphere and has a provable convergence guarantees.

Blind 1-bit CS algorithms is an ongoing research direction which has attracted more and more attentions in recent years [21,22].

In the next subsection, we exploit the compressible feature of sensor data and propose a novel blind 1-bit CS reconstruction algorithm based on BIHT.

### B. Our framework

In WSNs, the reading $x \in R^n$ each sensor node collects has intra-temporal correlation and thus can be sparsified under certain transform basis. Now, the projections for sensor reading is

$$b = sign(\Phi x) = sign(\Phi \Psi s) = sign(As) \quad (6)$$

where $b \in \{1,-1\}^m$, $\Phi$ is a given $m \times n$ Gaussian random matrix, $x = \Psi s$, $A = \Phi \times \Psi$, and $\Psi \in R^{n \times n}$ is an orthogonal transform matrix.

Since the scale information is lost, we propose that each sensor node needs to send extra norm value of $x$ to base station. Algorithm 1 gives out our general reconstruction framework running on the side of base station.

| Algorithm 1 Reconstruction Framework |
|---|
| 1) **Inputs**: vector of 1-bit measurements $b \in \{1,-1\}^m$ computing from equation (6), Gaussian random matrix $\Phi \in R^{m \times n}$, orthogonal transformation matrix $\Psi \in R^{n \times n}$ and the norm value of original sensor reading $norm\_x = \|x\|_2$. <br> 2) $\hat{s}$=**blind_1_bit_cs**($A$, $b$), here blind_1_bit_cs function can be any blind CS reconstruction algorithm which needs no |

prior sparse level information.

3) Get the estimate sensor readings $\acute{x} = (\Psi^T * \acute{s}) \times norm\_x$, here $\Psi^T$ is the transpose of $\Psi$, and $\Psi^T \times \Psi = I$.

Observing that only a small portion of coefficients dominate the energy of s, and those coefficients almost centralize at the beginning of s (see figure 1 and figure 2 in the next experiment section as examples), we propose a blind 1-bit CS algorithm based on BIHT[19]. We call our algorithm BBIHT, and the details are described in algorithm 2.

**Algorithm 2 BBIHT Reconstruction Algorithm**
1) **Initialization**: $max\_K = d \times n$ ( $0 < d < 1$ ), $S = zeros(n, max\_K)$, stop_var=0.01, best_k = 1
2) for k=1:1:max_K
3) $\acute{s}$ =BIHT(A,b,k); //using BIHT as the 1-bit CS reconstruction algorithm with given sparse level k
4) S(:,k) = $\acute{s}$;
5) end
6) for i=1:1:max_K
7) tmp = S(i,i:max_K);
8) if var(tmp) > stop_var //find the best result
9) best_k = i-1;
10) end
11) end
12) return S(:, best_k); // the best reconstructed signal

The key idea in Algorithm 2 is computing the variance of the rows of BIHT solutions giving different sparse level, and trying to find the best sparse level (from step 7 and 10 in algorithm 2). If the variance is larger than threshold *stop_var* which is a relative small value, we can tell that the related sparse level is not the best sparse level and then use previous sparse level as the best sparse level.

In the next section, we can find that BBHIT works very well in real WSN datasets.

### III. EXPERIMENTAL RESULTS

In this section, we run several numerical experiments on two real WSN datasets to demonstrate the effectiveness of our algorithm. The first dataset is a real world trace from Intel Berkeley Research lab and we choose the temperature information collected by sensor node with number 1 on March 2$^{nd}$, 2004 [23]. Another dataset is the temperature data collected in the Pacific Sea st (7.0N, 180W) on March 29, 2008 [24]. All experiments were performed using MATLAB on the same computer.

#### A. Sparsity of the WSN datasets

Figure 1 and figure 2 are the DCT coefficients of the two datasets, respectively. As we can see from the results, few coefficients dominate the energy, and most of coefficients are nearly zero. In fact, the first 4 DCT coefficients of Pacific Sea dataset occupy more than 99.97% energy. Thus, the two datasets are sparse under DCT transform and can be compressed using CS and 1-bit CS.

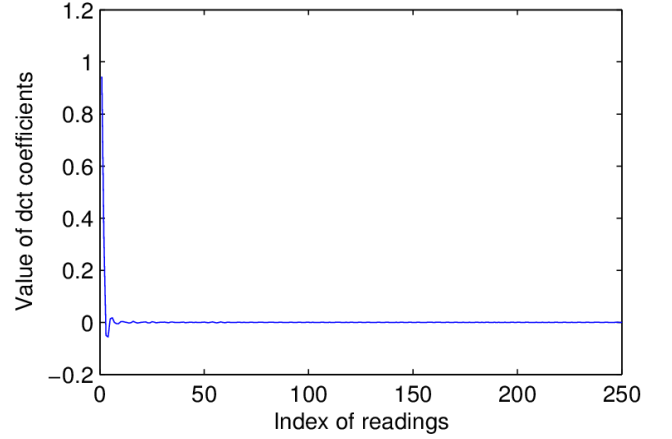

Fig. 1. DCT coefficients of Pacific Sea dataset. N=250.

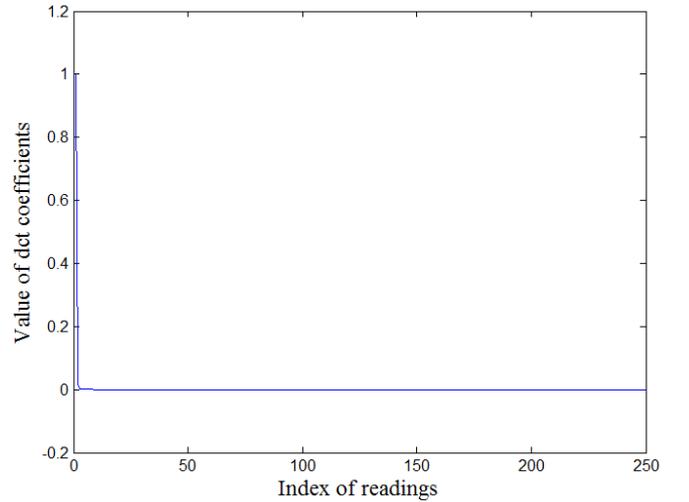

Fig. 2. DCT coefficients of Intel Berkeley Research lab dataset. N=250.

#### B. Reconstruction experiments

The setup for our experiments is as follows. The elements of matrix $\Phi \in R^{m \times n}$ follow i.i.d Gaussian distribution with zero mean and 1/m variance. $\Psi \in R^{n \times n}$ is a DCT orthogonal transform matrix. $d$ and *stop_var* in BBHIT are set to 0.1 and 0.01 respectively.

We performed 200 trials, and in each trial we fixed the length of reading vector $x$ as 250, and varied m from 25 to 500. The sign measurements are computed as in (6). We obtain estimated $\acute{x}$ using RSS [20], 1-bit FPC [18] and our BBHIT introduced in this paper, respectively.

The average signal-to-noise ratio (SNR) for each original reading $x$ with respect to estimated $\acute{x}$ was computed. Here

SNR is denoted by
$$20\log_{10}(\|x\|_2 / \|x - x'\|_2).$$

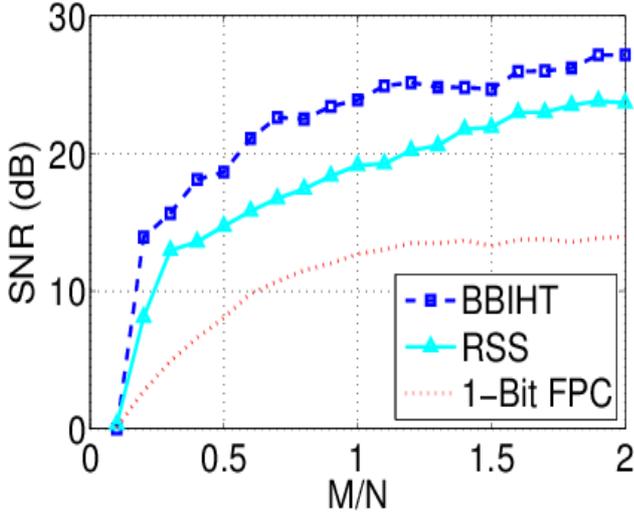

Fig. 3. SNR for Pacific Sea dataset. N=250. M varied from 25 to 500.

Figure 3 and figure 4 present the SNR values for Pacific Sea dataset and Berkeley Research lab dataset using RSS, 1-bit FPC and BBIHT algorithms respectively. The results demonstrate that BBIHT algorithm can achieve relatively higher accuracy than RSS and 1-bit FPC algorithms, especially for the Pacific Sea dataset which is sparser than the other dataset.

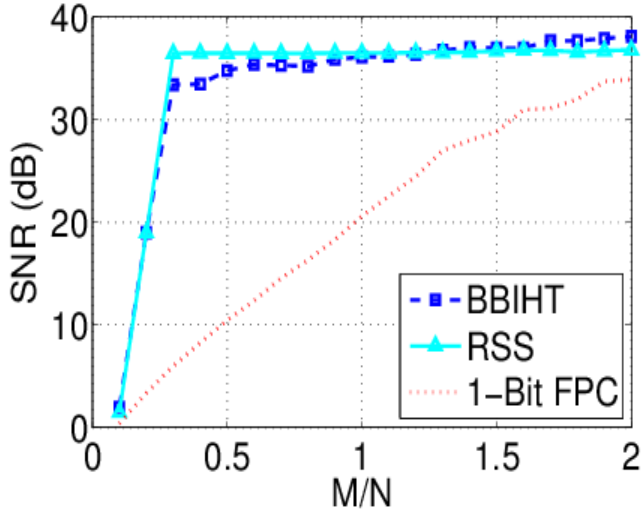

Fig. 4. SNR for Berkeley Research lab dataset. N=250, M varied from 25 to 500.

When M/N equal 1, figure 5 and figure 6 show the results of reconstructed readings and original readings for the two datasets respectively. As we can see, the constructed readings fit the original readings well.

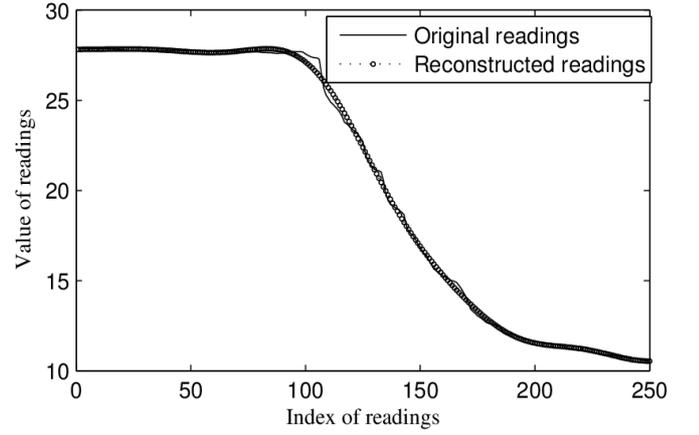

Fig. 5. Original readings Vs reconstructed readings for Pacific Sea dataset. N=250. M/N=1.

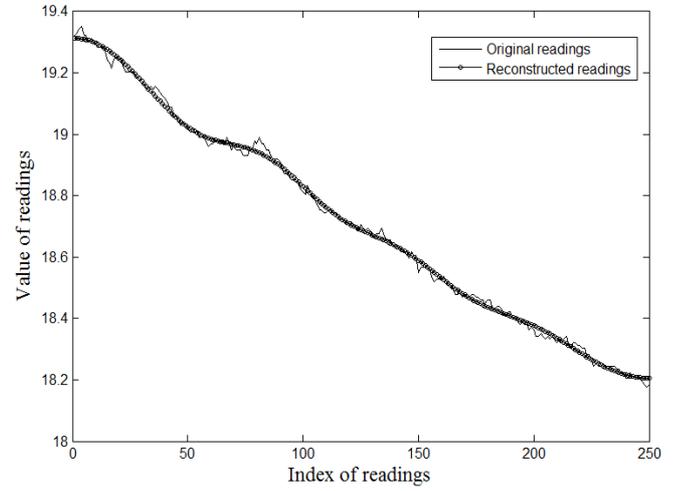

Fig. 6. Original readings Vs reconstructed readings for Intel Berkeley Research lab dataset. N=250. M/N=1.

### C. Comparison of compression ratio

Usually, sensor reading in WSN is IEEE single-precision floating point value, in order to present the traditional CS projection value 24 bits are needed. For example, the value of CS projection for Pacific Sea dataset lies between -914.8860 and 1.1418e+03, and 16 bits are not enough to store the projection. Thus, the compression ratio of traditional CS is
$$m \times 24 / (n \times 24) = m / n.$$

In CDG scheme the best compression ratio is 0.1 (section 5.1 in [5]).

In our BBIHT schema, it only need 1 bit to store each projection, and 24 bits to store the extra norm value $norm\_x$ in algorithm 1. Thus, the compression ratio of BBIHT in our experiment is
$$(m \times 1 + 24) / (n \times 24) = (m/24 + 1) / n.$$

When m>1 which is always the case, BBIHT compression ratio $(m/24 + 1)/n$ is much less than $m/n$. Hence,

BBIHT significantly reduces the communication overhead than traditional CS methods, and can still achieves a reasonable reconstruction accuracy. For example, as depicted in the figure 3 and figure 5, we can see that the SNR of Pacific Sea dataset is higher than 20db and the accuracy is enough when M/N = 1. At this case, the BBIHT compression ratio is $(250/24+1)/250 = 0.046$.

### D. Average CPU time

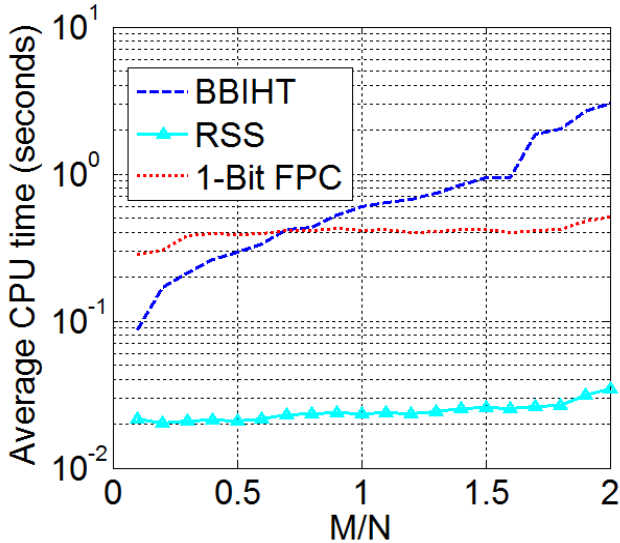

Fig. 7. Average CPU time for Pacific Sea lab dataset. N=250, M varied from 25 to 500.

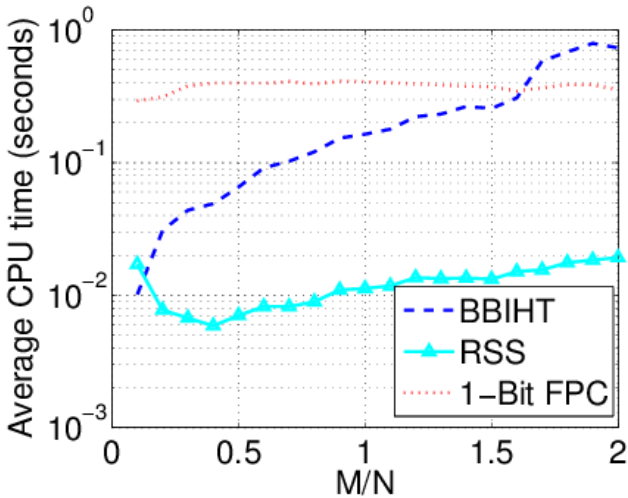

Fig. 8. Average CPU time for Intel Berkeley Research lab dataset. N=250, M varied from 25 to 500.

Figure 7 and figure 8 are the average CPU time of the three blind 1-bit CS algorithms for the two datasets respectively. We can see that the average CPU time of our BBIHT is below 1 second when M/N is less than 1. Also, note that we can adjust $d$ in algorithm 2 to balance the computation overhead and SNR accuracy.

## IV. CONCLUSION AND FUTURE WORK

In this paper, we have first presented a framework and a new blind 1-bit CS algorithm for the data gathering in WSNs. Experiments on two real world sensor datasets reveal our preliminary but interesting results, that is we can achieve much more compression ratio which means less communication overhead, and the reconstruction accuracy is still reasonable and acceptable SNR.

Theory analysis for BBIHT on compressible data and more experiments on other kind of real sensor datasets with different tuning parameters are remaining as our future work.

## V. ACKNOWLEDGEMENT

We would like to thank Zaiwen Wen for sharing their implementation of the original RSS algorithm[20].